\documentclass[lettersize,journal]{IEEEtran}
\usepackage[caption=false,font=normalsize,labelfont=sf,textfont=sf]{subfig}
\usepackage{textcomp}
\usepackage{stfloats}
\usepackage{url}
\usepackage{verbatim}
\usepackage{tabularray}
\usepackage{subfig}
\usepackage{float} 
\usepackage{multirow}
\usepackage{textcomp}
\usepackage{xcolor}
\usepackage{tablefootnote}
\usepackage{tabto}
\usepackage{breqn}
\usepackage{lipsum}
\usepackage{colortbl}
\usepackage{hhline}
\usepackage[super]{nth}
\usepackage{nicematrix,enumitem}
\usepackage[para]{threeparttable}
\usepackage[noend]{algpseudocode}
\usepackage{algorithm}
\usepackage{makecell}
\usepackage{setspace}
\usepackage{listings}
\usepackage{color}
\begin{document}

\title{KALAM: toolKit for Automating high-Level synthesis of Analog computing systeMs}

\author{Ankita Nandi,\IEEEmembership{}
        Krishil Gandhi,\IEEEmembership{}
        Mahendra Pratap Singh,\IEEEmembership{}
		Shantanu Chakrabartty,\IEEEmembership{}
		and~Chetan Singh Thakur\IEEEmembership{ }
\thanks{Ankita Nandi and Chetan Singh Thakur are with the NeuRonICS Lab, Department of Electronic Systems Engineering, Indian Institute of Science, Bengaluru 560012, India (E-mail: ankitanandi@iisc.ac.in, csthakur@iisc.ac.in). Krishil Gandhi and Mahendra Pratap Singh are students of Sardar Vallabbhai National Institute of Technology and Manipal Institute of Technology, respectively, interning at the NeuRonICS lab. Shantanu Chakrabartty is associated with the Department of Electrical \& Systems Engineering, Washington University in St. Louis, St. Louis, MO 63130 USA (E-mail: shantanu@wustl.edu). 
}}


\maketitle

\begin{abstract}

Diverse computing paradigms have emerged to meet the growing needs  for intelligent energy-efficient systems. The Margin Propagation (MP) framework, being one such initiative in the analog computing domain, stands out due to its scalability across biasing conditions, temperatures, and diminishing process technology nodes. However, the lack of digital-like automation tools for designing analog systems (including that of MP analog) hinders their adoption for designing large systems. The inherent scalability and modularity of MP systems present a unique opportunity in this regard. This paper introduces KALAM (toolKit for Automating high-Level synthesis of Analog computing systeMs), which leverages factor graphs as the foundational paradigm for synthesizing MP-based analog computing systems. Factor graphs are the basis of various signal processing tasks and, when coupled with MP, can be used to design scalable and energy-efficient analog signal processors. Using Python scripting language, the KALAM automation flow translates an input factor graph to its equivalent SPICE-compatible circuit netlist that can be used to validate the intended functionality. KALAM also allows the integration of design optimization strategies such as precision tuning, variable elimination, and mathematical simplification. We demonstrate KALAM's versatility for tasks such as Bayesian inference, Low-Density Parity Check (LDPC) decoding, and Artificial Neural Networks (ANN). Simulation results of the netlists align closely with software implementations, affirming the efficacy of our proposed automation tool.

\end{abstract}


\newcommand{\rom}[1]{\uppercase\expandafter{\romannumeral #1\relax}}

\maketitle

\section{Introduction}

Rising energy and performance demands have curated the need for advancements in analog hardware to create smarter machine interfaces with data compression and the development of new computing paradigms~\cite{srcDecadalPlan}.
In this regard, the Margin Propagation (MP) circuits offer a promising solution as a versatile analog computing framework, successfully demonstrated on hardware for signal processing (SP) and machine learning (MP) processors~\cite{ldpc_minggu,pk1, pk2}. MP allows energy efficient, accuracy tunable modular implementation of any non-linear monotonic function using piece-wise linear splines. MP analog designs can be pre-characterized and are scalable across temperature, biasing regimes, and process technology nodes, thus making MP a favorable analog standard cell~\cite{aryabhat}. 

However, despite its versatility and efficient implementations of ML and SP processors, MP analog systems lack popularity due to the absence of a digital-like automation framework for quick design synthesis and validation. Hence, in this work, we leverage the advantages of MP and propose an automation tool, KALAM. KALAM utilizes factor graphs to represent the system to be implemented while generating a SPICE-compatible netlist for design validation. Factor graphs can model various signal processing tasks, such as Kalman filters and stochastic models~\cite{factor_graphs}. When combined with MP, they create a powerful platform for designing analog signal processors. 
In this paper, we demonstrate KALAM's capabilities by designing Bayesian networks, LDPC decoders, and an ANN. The key contributions of this paper are:
\begin{itemize}[leftmargin=*] 
    \item An automation tool to synthesize MP analog computing circuits using input factor graphs to produce SPICE netlists.

    \item Implementation of Bayesian inference for different networks with varying nodes and features. The accuracy of KALAM-generated circuits corresponds to software accuracy. 

    \item Implementation of an MP-based LDPC decoder proposed in~\cite{ldpc_minggu} for $32-$bit, $64-$bit and $96-$bit codelengths using KALAM. The netlist-derived Bit and Frame Error Rate plots strongly correlate with software results.

    \item Implementation of an ANN on the IRIS dataset. The netlist-derived accuracy strongly correlates with software results.
\end{itemize}

The rest of the paper is organized as follows: Section~\ref{background} delves into the factor graph-based systems to be implemented. Section~\ref{proposed_synthesis} describes the proposed tool KALAM, and Section~\ref{results} delves into the results. A detailed discussion is presented in Section~\ref{discussion}, and Section~\ref{conclusion} concludes the work.

\section{Factor Graphs}\label{background}
Factor graphs are bipartite graphs consisting of two distinct sets of nodes. Variable nodes denoted by $\mathcal{V} = \{v_1, v_2, \cdots, v_N\}$, represent the random variables in the system. Factor nodes denoted as $\mathcal{F} = \{f_1, f_2, \cdots, f_M\}$, represent functions that constrain the probability distribution over the connected random variables. Each factor node (representing a function) $f_i$ connects to a set of variable nodes ($\mathcal{X}_i$) by edges.
The joint probability distribution of the factor graph is given by~\eqref{prob_dist}, where $Z$ is a normalization constant that ensures the probability distribution sums (or integrates) to 1.
\begin{equation}
p(v_1, v_2, \cdots, v_N) = \frac{1}{Z} \prod_{i=1}^{M} f_i({\mathcal{X}_i}) \label{prob_dist}
\end{equation}
This factorization allows efficient inference algorithms to be applied to the model. In this work, we discuss the following examples of factor graphs:

\begin{figure*}
\centering
\includegraphics[width=0.95\linewidth]{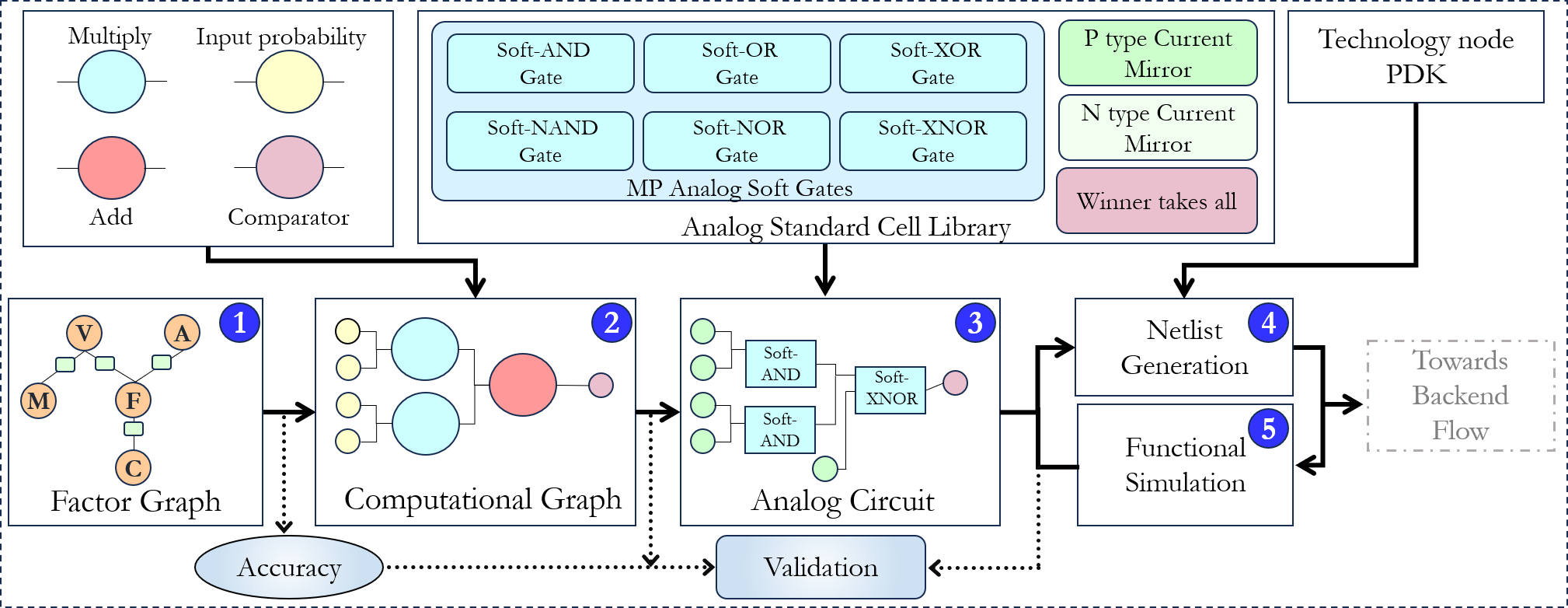}
\caption{This figure shows the proposed flow of KALAM for synthesizing analog MP-based systems. The associated figures in each stage refer to the design of a Bayesian network for illustration. \textit{Stage}~$1$ of this figure shows the factor graph representation to be synthesized. This factor graph is translated into a computational graph in \textit{Stage}~$2$, which is then mapped onto analog circuits in \textit{Stage}~$3$. The MP-based analog standard cell library is input at \textit{Stage}~$3$. Finally, the corresponding netlist is then generated in \textit{Stage}~$4$ and validated for accuracy through functional simulation in \textit{Stage}~$5$.}
\label{synthesis_flow}
\end{figure*}

\subsubsection{Bayesian Networks (BN) as Factor Graphs}

Bayesian networks are represented as Directed Acyclic Graphs (DAGs) that reflect the conditional probability distributions of each node $v_i$ given its parents as shown in~\eqref{prob_dist}, such that $ f_i({\mathcal{X}_i}) \equiv p(v_i|parents(v_i))$ and $Z=1$. 
Each factor node can be implemented using probabilistic soft gates, which in this case are MP analog soft gates~\cite{vlsid}. 

\subsubsection{Message Passing Decoders as Factor Graphs}
Probabilistic message passing between factor and variable nodes uses the Sum-Product Algorithm (SPA)~\cite{ldpc_minggu}, which is an iterative decoding mechanism for LDPC codes. 
The primary functions governing $v_i$ and $f_j$  are defined as under~\cite{vigoda_thesis}:
\begin{itemize}[leftmargin=*] 
  \item Message from $v_i$ to $f_j$ is given by~\eqref{eq:var_to_check_msg} where ($\mathcal{N}_{v \rightarrow f}$) shows all connected factor nodes to $v_i$:
  \begin{equation}
  m_{v_i \rightarrow f_j} = \prod_{f_k \in \mathcal{N}_{v \rightarrow f} \forall k\neq j}  m_{f_k \rightarrow v_i} \label{eq:var_to_check_msg}
  \end{equation}

  \item Message from $f_j$ to $v_i$ such that all connected variable nodes are written as ($\mathcal{X}_{f \rightarrow v}$), can be calculated as:
  \begin{equation}
  m_{f_j \rightarrow v_i} = \prod_{v_k \in \mathcal{X}_{{f \rightarrow v}} \forall k\neq i} m_{v_k \rightarrow f_j} \label{eq:check_to_var_msg}
  \end{equation}
\end{itemize}
In this work, we leverage KALAM to implement the MP-based LDPC decoder proposed in~\cite{ldpc_minggu}.

\subsubsection{Artificial Neural Network (ANN) as Factor Graphs}

In an ANN, the neurons typically represent the variable nodes. The factor nodes correspond to activation functions (e.g., sigmoid, ReLU) applied to the weighted sum of inputs to a neuron (also known as the Multiply-Accumulate (MAC) operation). In this work, we demonstrate KALAM for ANN inference by using the MP MAC and MP activation functions proposed in~\cite{pk1}.

The focus of this manuscript is to propose an automation framework for the design of MP-based analog processors like inference engines and decoders. Hence we refrain from discussing upon the merits of using MP designs.

\section{KALAM: Proposed Automation Framework}\label{proposed_synthesis}
This section presents the proposed automation flow, segmented into five stages as shown in Fig.~\ref{synthesis_flow}. We also present an example Bayesian network implementation towards the end of this section.

\subsection{Factor Graph Generation: Defining the system}
The process begins by defining the system as an equivalent factor graph. KALAM offers two functionalities:
\begin{itemize}[leftmargin=*] 
	\item \textbf{User-Provided Factor Graph}: If a factor graph already exists, the user can directly input it into the framework. For example, when designing LDPC decoders, we can leverage the factor graph given by the code's parity-check matrix.
	\item \textbf{Factor Graph Construction}: Alternatively, KALAM can construct a factor graph based on the user's input, such as a Bayesian network represented as a DAG.
\end{itemize}

\subsection{Computational Graph Generation: Mapping Operations}

Building upon the factor graph from \textit{Stage}~$1$ in Fig.~\ref{synthesis_flow}, \textit{Stage}~$2$ generates the computational graph. This graph depicts the crucial mathematical operations required at each factor node by using the following flow:

\begin{itemize}[leftmargin=*]
    \item \textbf{Analyzing the factor graph} to determine the number of inputs for each operation and the respective connections.
    \item \textbf{Computational optimization} eliminates redundant variables using variable elimination algorithms. It also implements mathematical and boolean simplifications if any.
    \item \textbf{Graph generation} by Graphviz Python package~\cite{graphviz} and \textbf{validation} against software implementation. 
\end{itemize} 

\begin{figure*}[t]
    \vspace{-1.5em}
    \centering
    \includegraphics[width=0.99\linewidth]{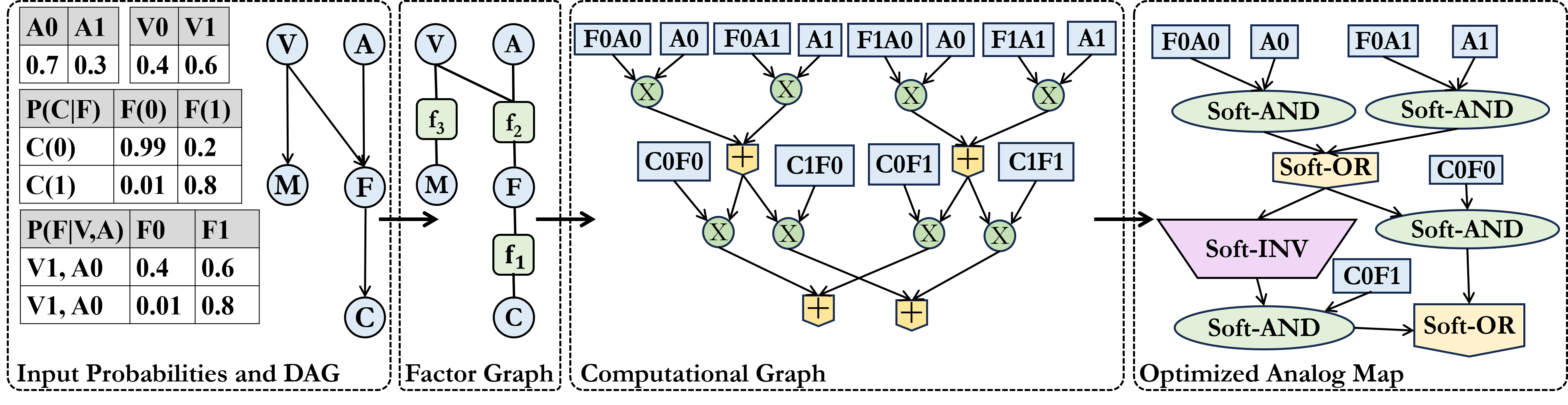}
    \caption{This figure is an example illustration of the different synthesis stages for a Bayesian DAG using its input probabilities. It is first translated into a factor graph. A computational graph is then generated with respect to the computations of each factor node after variable elimination. Optimization strategies such as Boolean simplification and precision tuning are employed before generating the final optimized analog map.}
    \label{case_study}
\end{figure*}

\begin{figure*}[t]

 \centering
\includegraphics[width=0.85\linewidth]{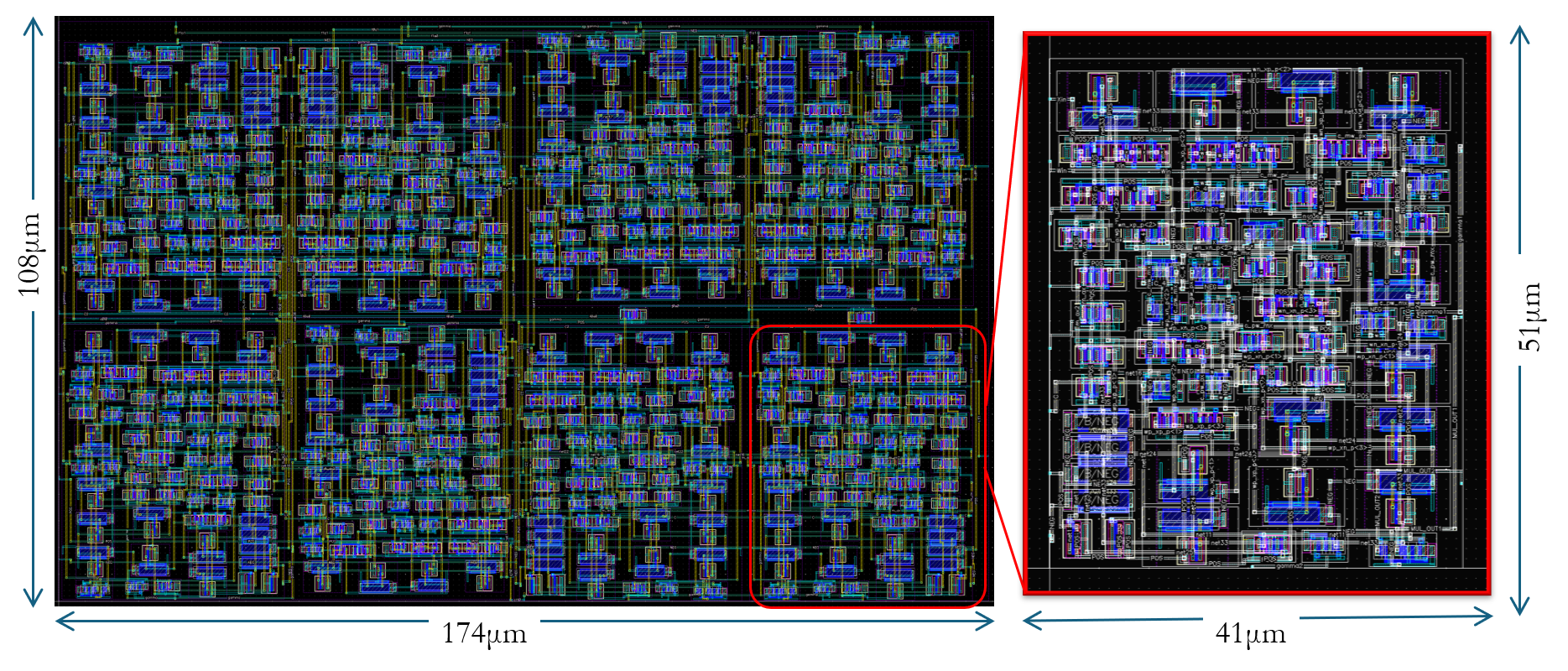}
 \caption{Layout for the KALAM-generated netlist corresponding to the Computational Graph in Fig.~\ref{case_study} consisting of eight Soft-AND gates (each mapped to a multiplier), using the place-and-route options offered by Cadence Layout GXL. The inset figure zooms into the layout of a single Soft-AND gate.}
 \label{layout}
 \end{figure*}

\subsection{Analog Circuit Mapping: From Graph to Circuits}

\textit{Stage}~$3$ translates the computations in the computational graph into practical analog circuits. This involves:
\begin{itemize}[leftmargin=*]
	\item \textbf{Optimizing computations} with corresponding standard cells from a pre-defined analog standard cell library implemented for varying accuracy, precision, power, and area budget. This is done by implementing the behavioural model design based on the computational graph for varying splines. Increasing splines increases the accuracy but  a penatlty of area overhead is to be paid~\cite{pk2}. Given that MP is modular, projecting the area overhead is feasible. A power-delay projection is made based on the operation regime. The weak inversion regime serves as a power saver mode while the strong inversion regime is the high performance mode~\cite{aryabhat}.
    \item \textbf{Functionally validating} this stage using software tools.
\end{itemize}

\subsection{Netlist Generation: Building the circuit}

The analog circuit map from \textit{Stage}~$3$ serves as the blueprint for netlist generation. Netlist contains the components and their interconnections, which are essential for circuit simulation. KALAM approaches netlist generation by:
\begin{itemize}[leftmargin=*]
	\item \textbf{Pre-defined Standard cells:} Synthesis of the standard cells involves computational mapping and transistor sizing using established techniques~\cite{vlsid, gmid_murman}.
	\item \textbf{Final netlist} is generated based on the analog circuit map by instantiating the implemented standard cells from the library using Python scripts for subsequent simulation.
\end{itemize}

\begin{table}[t]
\centering
\caption{Accuracy and Runtime of different Bayesian networks}
\label{accuracy_table}
\resizebox{0.9\columnwidth}{!}{%
\begin{tabular}{|c|c|c|c|c|}
\hline
{\cellcolor[HTML]{ECF4FF}\textbf{Model}} & {\cellcolor[HTML]{ECF4FF}\begin{tabular}[c]{@{}c@{}}$\#$\textbf{Variable} \\\textbf{Nodes}\end{tabular} }   & {\cellcolor[HTML]{ECF4FF}\begin{tabular}[c]{@{}c@{}}\textbf{Software} \\\textbf{Accuracy} (\%)\end{tabular} } & {\cellcolor[HTML]{ECF4FF}\begin{tabular}[c]{@{}c@{}c@{}}\textbf{Netlist} \\\textbf{Accuracy (\%)}\end{tabular} }& {\cellcolor[HTML]{ECF4FF}\begin{tabular}[c]{@{}c@{}}\textbf{Synthesis} \\\textbf{Runtime (ms)}\end{tabular} } \\ \hline
Airplane   & $22$  & $82.38$  & $82.27$   & $6.1$   \\ \hline
Titanic   & $8$   & $77.99$   & $77.91$   & $5.8$   \\ \hline
Pistachio & $17$  & $77.12$  & $77.12$   & $5.9$  \\ \hline
Thyroid  & $15$   & $95.3$   & $95.3$    & $5.9$    \\ \hline
Drought  & $19$   & $97.93$  & $97.93$   & $6 $ \\ \hline
\end{tabular}}
\end{table}

\subsection{Functional Simulation and Validation}
The generated netlists undergo validation using SPICE tools. We leverage \textit{Cadence virtuoso} for visual inspection and \textit{Cadence Spectre} for functional validation of the designs. Further, similar to digital synthesis the layout can be generated by using this schematic on \textit{Cadence Layout GXL} through the automatic place and route features of the standard analog cells.

\subsection*{Example Design of a Bayesian network Using KALAM}
To demonstrate the capabilities of KALAM, we illustrate various stages of the tool using an example involving the design and analysis of a simple Bayesian network to infer prey catching~\cite{chetan_bayesian}, as shown in Fig.~\ref{case_study}. It starts by factorizing the probabilities to convert them to a factor graph. Thus, the factor nodes are $f_1 = p(C|F) $, $f_2 = p(F|A,V)$. It can be observed that $f_3 = p(M|V)$ is unrelated to node \textbf{C}. For a small network such as this, we can visually eliminate unrelated variable nodes, but for large networks, this is not feasible. Thus, we need the computational graph. For the next stage, factor nodes are unrolled to represent computations in the computational graph, which is, in turn, used to make the analog map. The analog map is then transformed into a SPICE netlist. For SPICE validation, the input magnitudes have been mapped as a multiple of $1 \mu A$ current for each input, which can be tuned based on the power budget. The final output of the BN, seen as the output current of the Soft-OR, was found to be $0.524 \mu A$, corresponding closely to the expected mathematical output $p(C=1/V=1) = 0.555$.

In Fig.~\ref{layout}, we use the netlist which is generated using the computational graph with eight multipliers and two adders (prior to the optimization of the analog map). These multipliers are mapped to a Soft-AND operation and the adders are showcased by addition of currents (by Kirchoff's current conservation). Thus the corresponding layout has eight Soft-AND gates as seen in Fig.~\ref{layout} with a zoomed inset of a single Soft-AND.

\section{Results}\label{results}
We evaluate the tool's efficacy by validating the functionality of the KALAM-generated Bayesian Networks, LDPC decoders, and ANN circuits.

 \begin{figure*}[t]
 \vspace{-0.5em}
 \centering
 \subfloat[]{\includegraphics[width=0.28\linewidth]{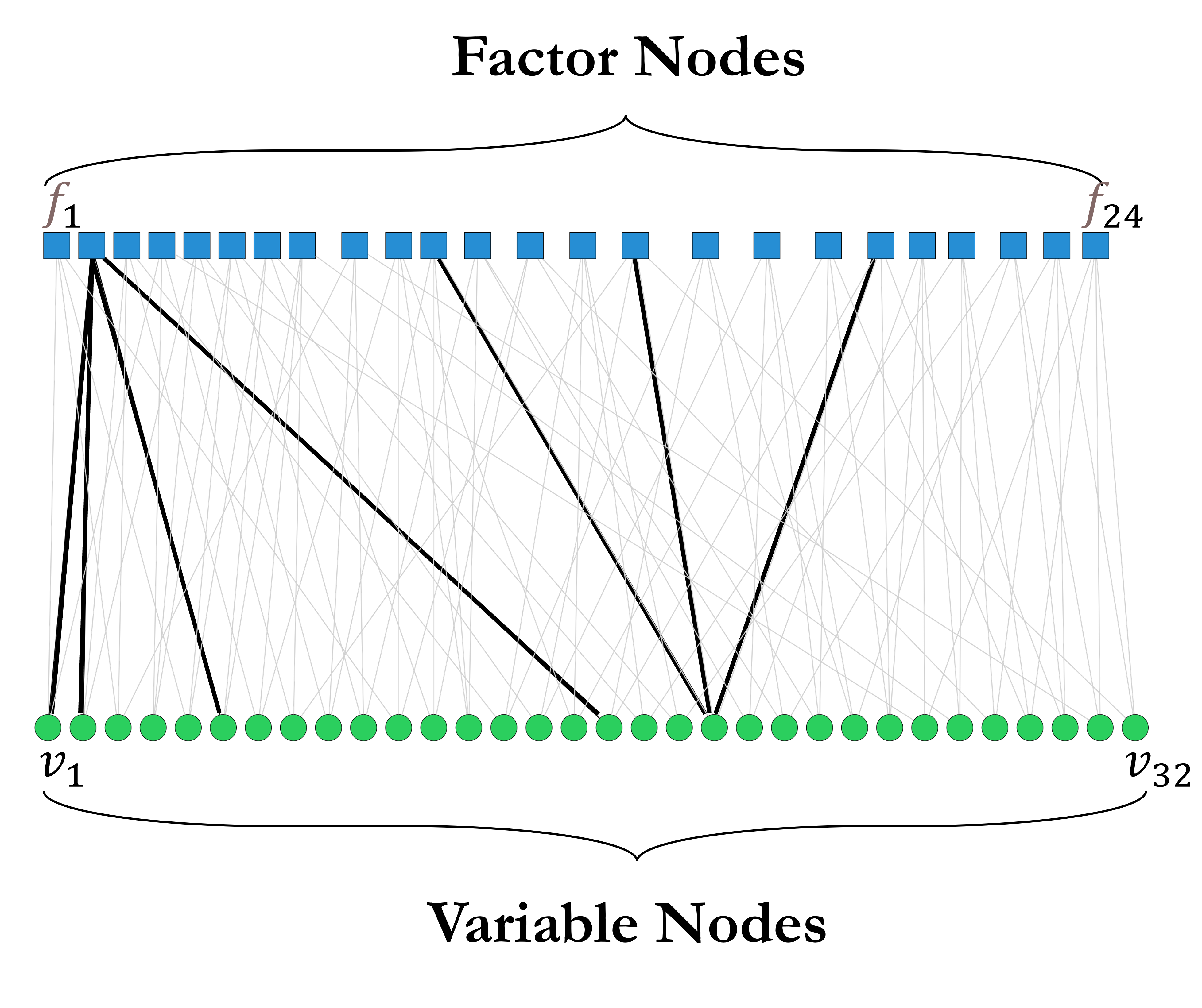}
 \label{tanner}}
 \subfloat[]{\includegraphics[width=0.27\linewidth]{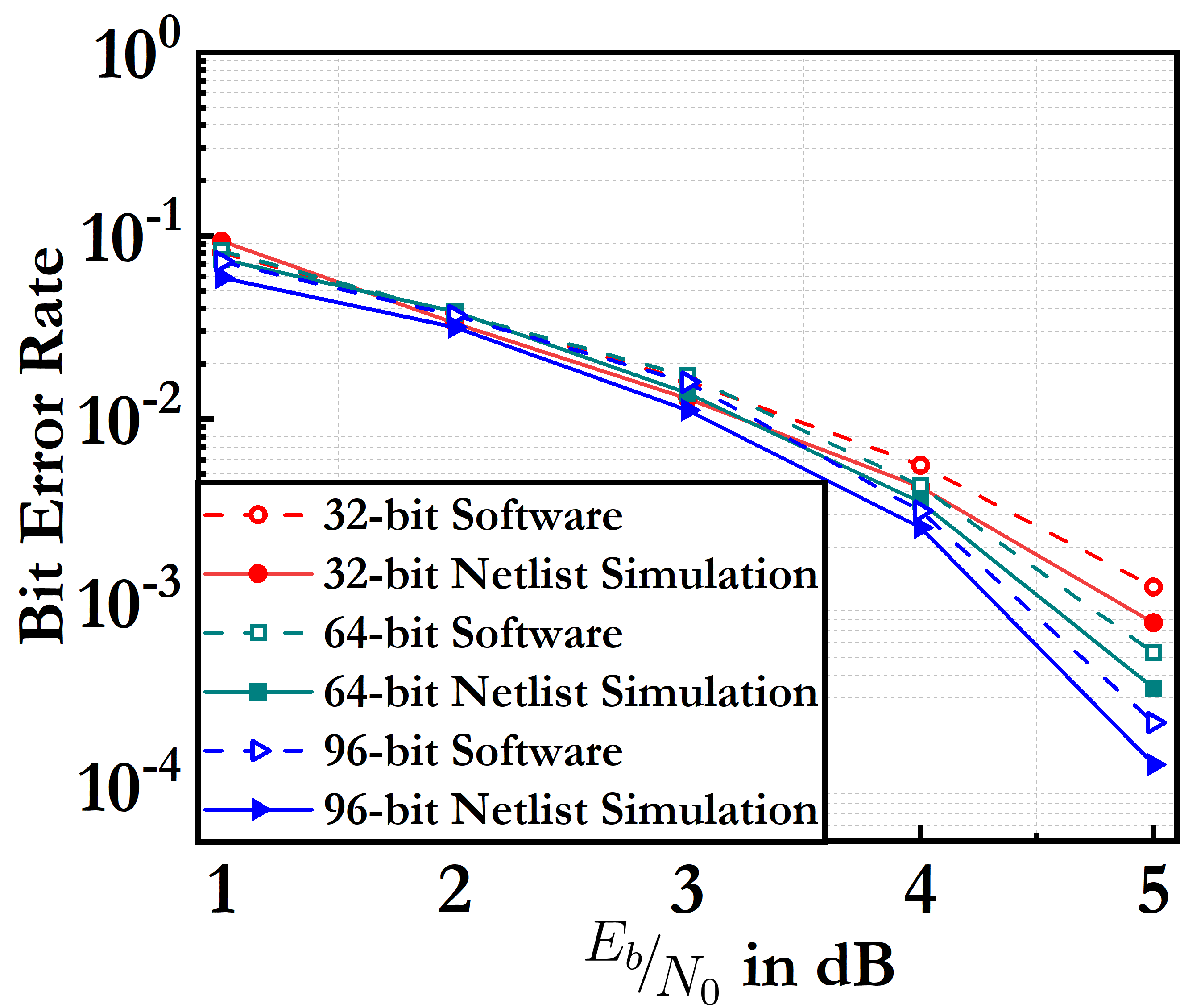}
 \label{ber}}
 \subfloat[]{\includegraphics[width=0.27\linewidth]{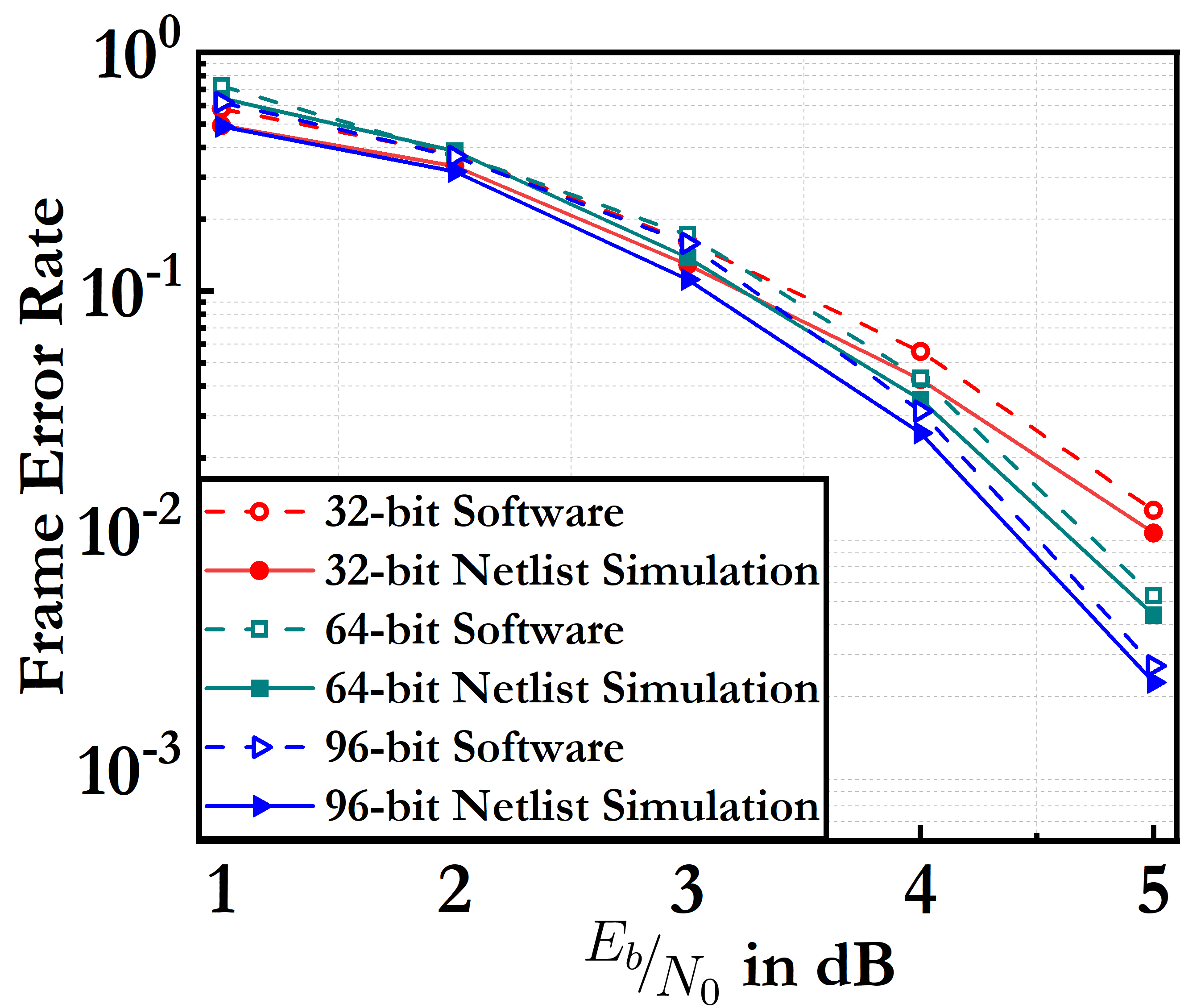}
 \label{fer}}
 \caption{\protect\subref{tanner} Tanner Graph for a (32,8) LDPC Decoder; \protect\subref{ber} Bit Error Rate (BER) and \protect\subref{fer} Frame Error Rate (FER) of the simulated netlist w.r.t. the software implementation for code lengths $32-$bit (synthesis runtime$=32 ms$), $64-$bit (synthesis runtime$=39 ms$) and $96-$bit (synthesis runtime$=51 ms$).}
 \label{ldpc_results}
 \end{figure*}

\begin{table*}[t]
\centering
\caption{Summary of Analog Design Automation Frameworks for Synthesis}
\label{comp_table}
\resizebox{\linewidth}{!}{
\begin{tabular}{|c|c|c|c|}

\hline
{\cellcolor[HTML]{ECF4FF}\textbf{Work}}     &{\cellcolor[HTML]{ECF4FF}\textbf{Methodology}}   & {\cellcolor[HTML]{ECF4FF}\textbf{Flow Automated}}      & {\cellcolor[HTML]{ECF4FF}\textbf{Example Design Automated}} \\ \hline
\cite{syn_1}                 & Franz Lisp                                                                    & \begin{tabular}[c]{@{}c@{}}Sizing and Netlist generation \\  \end{tabular}                                                                & Comparator Topology                                                       \\ \hline

\cite{syn_3}                 & Signal Flow Graph Transformation                                                        & \begin{tabular}[c]{@{}c@{}}Structural synthesis using Transfer function \end{tabular}                                                                      & Summing Circuit                                                           \\ \hline

\cite{syn_4}                 & Equation based Optimisation                  & Sizing    & CMOS Op-Amps                                                              \\ \hline
\cite{syn_5}                 & ODAE Model, VHDL-AMS, Java                   & \begin{tabular}[c]{@{}c@{}}Structural synthesis using VHDL-AMS\end{tabular}                                                                                                                                              & Oscillator                                                                \\ \hline
\cite{syn_6}                 & Adaptive Genetic Algorithm, C++                   & Topology    & OpAmp, Oscillator                                                         \\ \hline
\cite{syn_8} & \begin{tabular}[c]{@{}c@{}}Differential Evolution\end{tabular}      & \begin{tabular}[c]{@{}c@{}}Architectural selection and Sizing \end{tabular} & CMOS OpAmp  \\ \hline


LASER~\cite{syn_9}          & Tcl/Tk, C++, CPLEX                   & Netlist Synthesis and Layout & CMOS OpAmp \\ \hline

AutoCkt~\cite{autockt}        & Reinforcement learning                                                              & \begin{tabular}[c]{@{}c@{}}Sizing   \\ \end{tabular}  & \begin{tabular}[c]{@{}c@{}}CMOS Two-stage OpAmp, OTA, TIA\end{tabular} \\ \hline

\cite{dac_2022}             & Reinforcement learning: GCN                                                             & Sizing (Width \& Fingers)                                                      & \begin{tabular}[c]{@{}c@{}}Two-stage OpAmp; GaN RF PA  \end{tabular}                                                     \\ \hline

OPAMP-Generator\cite{syn_10}                & \begin{tabular}[c]{@{}c@{}}VGAE, Bayesian optimisation\end{tabular} & \begin{tabular}[c]{@{}c@{}c@{}}Sizing and Netlist generation \end{tabular}                                                               & \begin{tabular}[c]{@{}c@{}}CMOS OpAmp, Three-stage OpAmp\end{tabular}  \\ \hline

\cite{automation_tcasii}                & Aritificial Neural Networks, SIMSIDES~\cite{simsides} & \begin{tabular}[c]{@{}c@{}c@{}}Size and performance optimization \end{tabular}                                                               & \begin{tabular}[c]{@{}c@{}}$\Sigma\Delta$ Modulators, OTA\end{tabular}  \\ \hline

KALAM (This Work)          & \begin{tabular}[c]{@{}c@{}}Factor Graphs, Margin Propagation \end{tabular}       & Software to Netlist generation   & \begin{tabular}[c]{@{}c@{}}Bayesian networks,  LDPC Decoders \end{tabular} \\ \hline
\end{tabular}
}
\end{table*}

\subsection{Bayesian networks}

We used five different Bayesian networks from datasets found in Kaggle~\cite{kaggle}. The \textit{pgmpy}~\cite{ankan2015pgmpy} Python library is used to obtain the ground truth software accuracy. The runtime of KALAM to generate the netlists, the software accuracy and the inference accuracy obtained through DC analyses of these netlists on \textit{Cadence Spectre} is reported in Table~\ref{accuracy_table}.

\subsection{LDPC Decoder}

We implement $32-$bit regular LDPC code having a $4-$degree factor node and a $3-$degree variable node as presented in Fig.~\ref{tanner}. 
The design in~\cite{ldpc_minggu} comprises both MP and non-MP designs easily integrated by the tool. 
The runtime to generate the netlist is mentioned in Fig.~\ref{ldpc_results}. Further, we expand the $32-$bit code to $64-$bit and $96-$bit codelengths using the protograph technique~\cite{protograph} of code construction. The Bit and Frame Error plots in Fig.~\ref{ber} and Fig.~\ref{fer} show the netlist simulation. The netlist simulation results conform with the software accuracy, demonstrating the efficacy of our tool. 

\subsection{Artificial Neural Network (ANN)}
We demonstrate the design of an ANN using a network comprising of $4$ input nodes, $8$ hidden nodes, and $3$ output nodes. The network consists of a MAC operation and ReLU activation function. For a training accuracy of $92$\%, the software testing accuracy was found to be $90.5$\% for an IRIS dataset using MP computational modules. The simulation of the KALAM-generated netlist also provided a $90$\% accuracy.

\section{Discussion}\label{discussion}  

Table~\ref{accuracy_table} shows a negligible drop in the accuracy of the KALAM-generated netlists compared to their software counterpart for both Bayesian networks. Fig.~\ref{ber} and Fig.~\ref{fer} show that the bit and the frame error rates of the netlist simulations comply with the software implementation of the decoding algorithm. The difference in the software and the implemented SPICE results is an artifact also reported in the design in~\cite{ldpc_minggu}, which the authors of~\cite{ldpc_minggu} explain as the effect of continuous-time analog simulations. The similarity of the implemented results with the baseline design reported in~\cite{ldpc_minggu} is a testament to the tool's efficacy in implementing and integrating both MP and non-MP modules. The ANN has a software inference accuracy of $90.5\%$ while a netlist accuracy of $90\%$. The runtime to generate the netlists of all the designs implemented by KALAM is observed to be $< 1s$ for an \textit{Intel i7} processor, which is manifold compared to manual implementation considering the scale of the system being implemented.

Table~\ref{comp_table} shows that while significant efforts have been dedicated to optimizing fundamental analog circuits, the automation required for designing large scalable analog circuits is less explored. The ability of MP-based designs to be pre-characterized, modular, and robust like digital designs, makes them suitable for automating large analog computing systems.

\section{Conclusion}\label{conclusion}

In conclusion, this work introduces KALAM, an innovative automated synthesis flow for designing analog computing systems using the MP framework. KALAM handles the full design process, from high-level descriptions using factor graphs to SPICE-compatible netlist generation. It further supports the integration of non-MP modules, which is evidence that KALAM can be used for non-MP designs. Validation against custom designs demonstrates its potential for scalable and practical applications. Future versions will explore the synthesis of other graphical models and various machine-learning architectures for both MP and non-MP analog computing.

\section*{Acknowledgments}
The authors acknowledge the MoU between IISc Bangalore and Washington University in St. Louis for facilitating collaboration between the two institutions.

\bibliographystyle{IEEEtran}
\bibliography{ref}

\end{document}